\documentclass[conference]{IEEEtran}
\IEEEoverridecommandlockouts
\usepackage{cite}
\usepackage{microtype}
\usepackage{amsmath,amssymb,amsfonts,bm}
\usepackage{algorithmic}
\usepackage{graphicx}
\usepackage{textcomp}
\usepackage{xcolor}
\usepackage{listings}
\usepackage{dirtytalk}
\usepackage[charsperline=55]{jlcode}
\usepackage[hidelinks]{hyperref}
\usepackage{cleveref}
\def\BibTeX{{\rm B\kern-.05em{\sc i\kern-.025em b}\kern-.08em
    T\kern-.1667em\lower.7ex\hbox{E}\kern-.125emX}}
\begin{document}

\title{Productivity meets Performance: Julia on A64FX
}

\author{\IEEEauthorblockN{1\textsuperscript{st} Mosè Giordano}
\IEEEauthorblockA{\textit{Advanced Research Computing} \\
\textit{UCL}\\
London, United Kingdom \\
m.giordano@ucl.ac.uk}
\and
\IEEEauthorblockN{2\textsuperscript{nd} Milan Kl\"ower}
\IEEEauthorblockA{\textit{Atmospheric, Oceanic and Planetary Physics} \\
\textit{University of Oxford}\\
Oxford, United Kingdom \\
milan.kloewer@physics.ox.ac.uk }
\and
\IEEEauthorblockN{3\textsuperscript{rd} Valentin Churavy}
\IEEEauthorblockA{\textit{CSAIL, EECS} \\
\textit{Massachusetts Institute of Technology}\\
Cambridge, United States of America \\
vchuravy@mit.edu}
}

\maketitle

\begin{abstract}
The Fujitsu A64FX ARM-based processor is used in supercomputers such as Fugaku in Japan and Isambard 2 in the UK and provides an interesting combination of hardware features such as Scalable Vector Extension (SVE), and native support for reduced-precision floating-point arithmetic. The goal of this paper is to explore performance of the Julia programming language on the A64FX processor, with a particular focus on reduced precision.
Here, we present a performance study on \texttt{axpy} to verify the compilation pipeline, demonstrating that Julia can match the performance of tuned libraries.
Additionally, we investigate Message Passing Interface (MPI) scalability and throughput analysis
on Fugaku showing next to no significant overheads of Julia of its MPI interface.
To explore the usability of Julia to target various floating-point precisions, we present results of \texttt{ShallowWaters.jl}, a shallow water model that can be executed a various levels of precision.
Even for such complex applications, Julia's type-flexible programming paradigm offers both, productivity and performance.
\end{abstract}


\section{Introduction}
The Julia programming language~\cite{Bezanson2017-ca} is a dynamic programming language, with a focus on productivity and performance.
It has found increased adoption in numerical and scientific computing, data processing and analytics, differentiable programming and scientific machine learning.
Julia uses LLVM~\cite{Lattner2004-hw} as a compiler backend and supports multiple CPU architectures (x86, PPC, ARM) which includes Fujitsu's A64FX that powers the Fugaku supercomputer.
Based at the RIKEN Center for Computational Science (R-CCS) in Kobe, Japan, Fugaku is currently number 2 in the TOP500 ranking of the fastest supercomputers in the world. It has topped the list from its induction in June 2020 to June 2022~\cite{dongarra:fugaku}.
It is composed of 158\,976 Fujitsu A64FX FX1000 CPUs, making it the first ARM supercomputer to claim the highest position in TOP500.
Inter-node communication is powered by Tofu Interconnect D (TofuD), a proprietary technology developed by Fujitsu~\cite{ajima:tofud}. 

While 16-bit arithmetic are increasingly supported on modern hardware, 64-bit double-precision floating-point numbers (here called \texttt{Float64}, other formats likewise, referring respectively to the IEEE-754 standard formats~\cite{noauthor_2019-pu}) are still widely used in scientific computing as they largely remove the necessity of a detailed numerical analysis of rounding errors in most applications.
Smaller rounding errors and lower risk of under and overflows come at a cost of
computational performance as low-precision arithmetic is executed significantly faster on supporting hardware.
Also \texttt{Float32} is widely supported on various CPU architectures, and many GPUs support different 16-bit formats (\texttt{Float16}, \texttt{BFloat16}), but A64FX is the first modern CPU for high-performance computing (HPC) that supports \texttt{Float16} arithmetic.
Many mantissa bits in high-precision formats contain little to no information in applications with large uncertainties, such as deep learning~\cite{Wang2018,Kalamkar2019} or climates models~\cite{Klower2021a}.
In these examples, a higher performance from low-precision arithmetic could be exchanged for larger networks or higher resolution to increase the complexity of these models.
In theory, A64FX promises $4x$ performance increase of \texttt{Float16} over \texttt{Float64} in both memory and compute-bound applications through its 512-bit SVE vectorization unit~\cite{FUJITSU2020}.

This article is an experience report with Julia on Fujitsu A64FX, a processor architecture primarily used by supercomputers \cite{Odajima2020,Sato2020}. We focus on the performance of low-precision float arithmetic and evaluate the overheads of using Julia for MPI programs on Fugaku. It follows a section on ShallowWaters.jl, a shallow water model, that makes use of Julia's type-flexibility to be executable at various levels of precision.

\section{Support for low-precision floating-point arithmetic in Julia}
\label{sec:fp16_support}
Julia uses multiple dispatch which as a programming paradigm makes it easier to support new number formats. In the code snippet below we reproduces the type hierarchy of floating-point numbers,
starting with the abstract type \texttt{Number} and ending at the primitive type \texttt{Float16}.

\begin{jllisting}
abstract type Number end
abstract type Real <: Number end
abstract type AbstractFloat <: Number end
primitive type Float64 <: AbstractFloat 64 end
primitive type Float32 <: AbstractFloat 32 end
primitive type Float16 <: AbstractFloat 16 end
\end{jllisting}

This type-hierarchy comes into play when we look at the implementation of math routines, like the cube root function \texttt{cbrt}.

\begin{jllisting}
julia> methods(cbrt)
# 7 methods for generic function "cbrt":
[1] cbrt(x::Union{Float32, Float64}) in Base.Math at special/cbrt.jl:142
[2] cbrt(a::Float16) in Base.Math at special/cbrt.jl:150
[3] cbrt(x::BigFloat) in Base.MPFR at mpfr.jl:626
[4] cbrt(x::AbstractFloat) in Base.Math at special/cbrt.jl:34
[5] cbrt(x::Real) in Base.Math at math.jl:1352    
\end{jllisting}

Julia provides for \texttt{cbrt} several implementations that range from the specialized to the generic.
\texttt{Float32} and \texttt{Float64} share an implementation and \texttt{Float16} is separated. Julia then dynamically dispatches to the most specific method available for a given type at runtime.
This allows Julia to both provide general implementations and fast implementations that take advantage of the structure of the types.

Early versions of Julia supported \texttt{Float16} only as a storage format. Mathematical functions immediately promoted to \texttt{Float32} and no rounding was performed.
This changed in Julia v0.6\footnote{\url{https://github.com/JuliaLang/julia/pull/17297}.}, and since then operations converted their output back to \texttt{Float16}.
Yet this was fully done in software and no hardware support (even if available) was used. Since Julia v1.6\footnote{\url{https://github.com/JuliaLang/julia/pull/37510}.} 
the compiler lowers the \texttt{Float16} type to LLVM's \texttt{half} type. 

The default behavior of LLVM on x86 chips was to extend the precision of \texttt{half} operations to \texttt{float}, 
which is unsuitable for numerical implementations that need to return consistent results on software and hardware implementations.
GCC recognized this problem as well in version 12~\cite{noauthor_undated-nw}: \say{The default behavior for \texttt{FLT\_EVAL\_METHOD} is to keep the intermediate result of the operation as 32-bit precision. This may lead to inconsistent behavior between software emulation and AVX512-FP16 instructions}. In Julia we generally require that operations are numerically stable across different hardware platforms,
and thus for software \texttt{Float16} we insert rounding operations.
On hardware that support \texttt{Float16} natively we could use LLVM to directly lower to hardware instructions.
There is ongoing work detailed in \cref{subsec:better_compiler} to improve compiler
support for detecting hardware that supports \texttt{Float16}.
For the experiments in \cref{subsec:float16} we explicitly turn on this support\footnote{\url{https://github.com/JuliaLang/julia/issues/40216}.}.

\section{Type-flexibility and performance}

\subsection{Performance and Scalability on Fugaku}
\label{subsec:scalability}

We run benchmarks of Julia code on Fugaku to evaluate the efficiency of the code generated by LLVM for simple Julia functions, and the overhead of using the \texttt{MPI.jl} package for communications on a world-class supercomputer.
The results of all these benchmarks are publicly available at \url{https://github.com/giordano/julia-on-fugaku}, including the job scripts used to run the benchmarks and the code to produce the plots reported in the present section, along with the Julia package environments used, for full reproducibility.

\subsubsection{Level 1 BLAS routine}
\label{sec:blas-axpy}

The Basic Linear Algebra Subprograms (BLAS) is a prescription of low-level routines for performing common linear algebra operations, such as matrix and vector operations~\cite{lawson:blas}.
BLAS routines are classified in three levels: Level 1 includes vector and vector operations, Level 2 includes vector-matrix operations, Level 3 includes matrix-matrix operations.
There are several high-performance implementations of BLAS routines, with hardware vendors often providing highly optimised BLAS implementations for their own systems.
One of the most common Level 1 routines is \texttt{axpy} which represents the mathematical operations of multiplying a vector $\pmb{x}$ by a scalar $a$, adding it to a vector $\pmb{y}$ and storing the result back into $\pmb{y}$ (``\emph{a} times \emph{x} \emph{p}lus \emph{y}''):
\begin{equation}
    \label{eq:axpy}
    \pmb{y} \leftarrow a \pmb{x} + \pmb{y}.
\end{equation}

We implemented in Julia a generic single-threaded \texttt{axpy} function, which can take in input vectors of any arbitrary numerical Julia type:
\begin{jllisting}
function axpy!(a::T, x::Vector{T}, y::Vector{T}) where {T<:Number}
    @simd for i in eachindex(x, y)
        @inbounds y[i] = muladd(a, x[i], y[i])
   end
   return y
end
\end{jllisting}
The \texttt{@simd} macro suggests the compiler to enable Single Instruction Multiple Data (SIMD) instructions, and the macro \texttt{@inbounds} informs the compiler it is safe to skip bounds checks, as we automatically iterate over the existing indices of the vectors \texttt{x} and \texttt{y} with the \texttt{eachindex} function.
In \cref{sec:fp16_support} we note that Julia is currently supposed to widen \texttt{Float16} numbers to \texttt{Float32} numbers, but due to a bug in the LLVM pass\footnote{\url{https://github.com/JuliaLang/julia/issues/45881}.} vectors of \texttt{Float16} are not widened to vectors of \texttt{Float32}, thus this implementation of \texttt{axpy} retains hardware performance for 16-bit floating-point numbers as desired.

\begin{figure}
    \centering
    \includegraphics[width=\columnwidth]{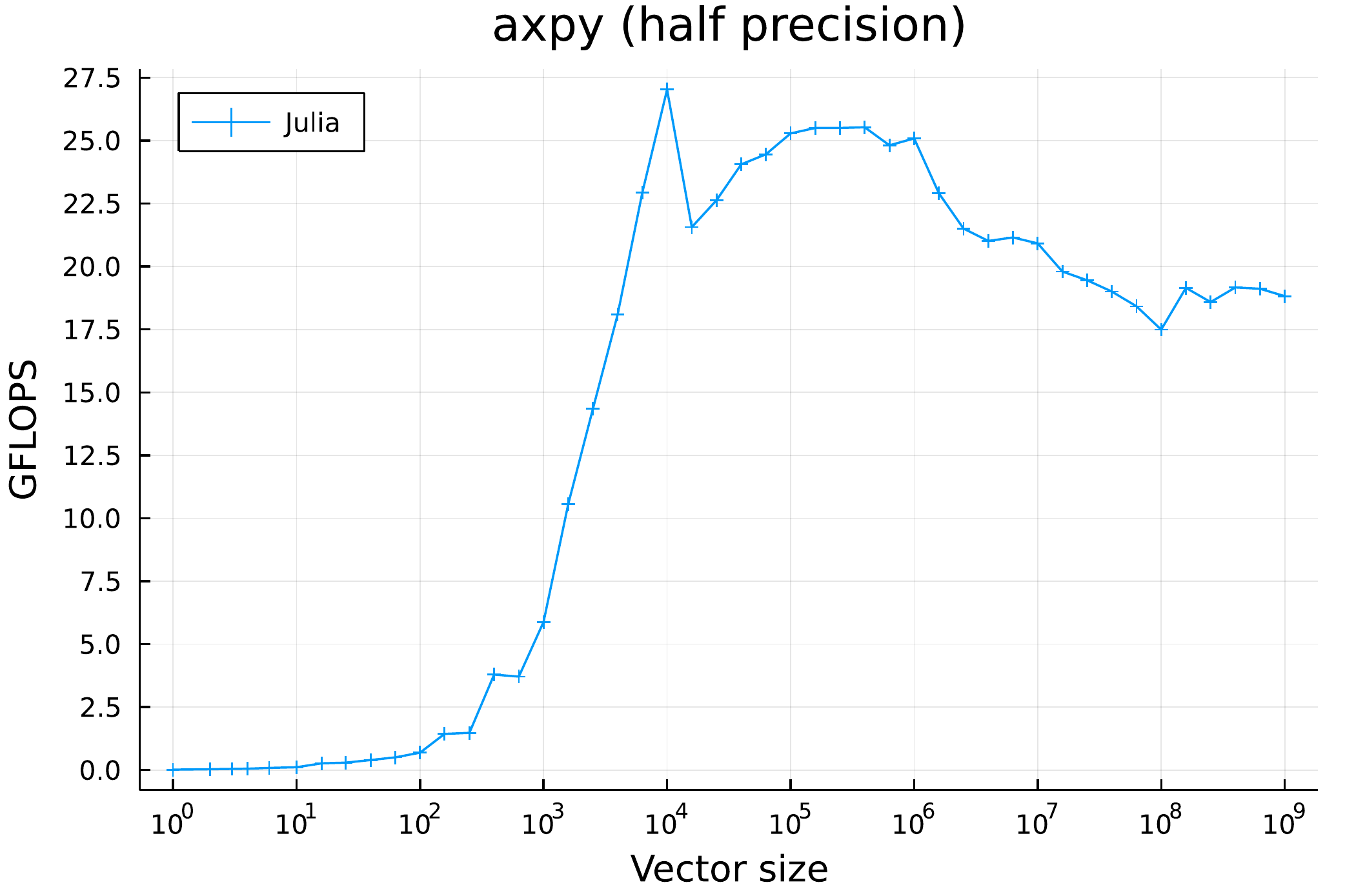} \\[1em]
    \includegraphics[width=\columnwidth]{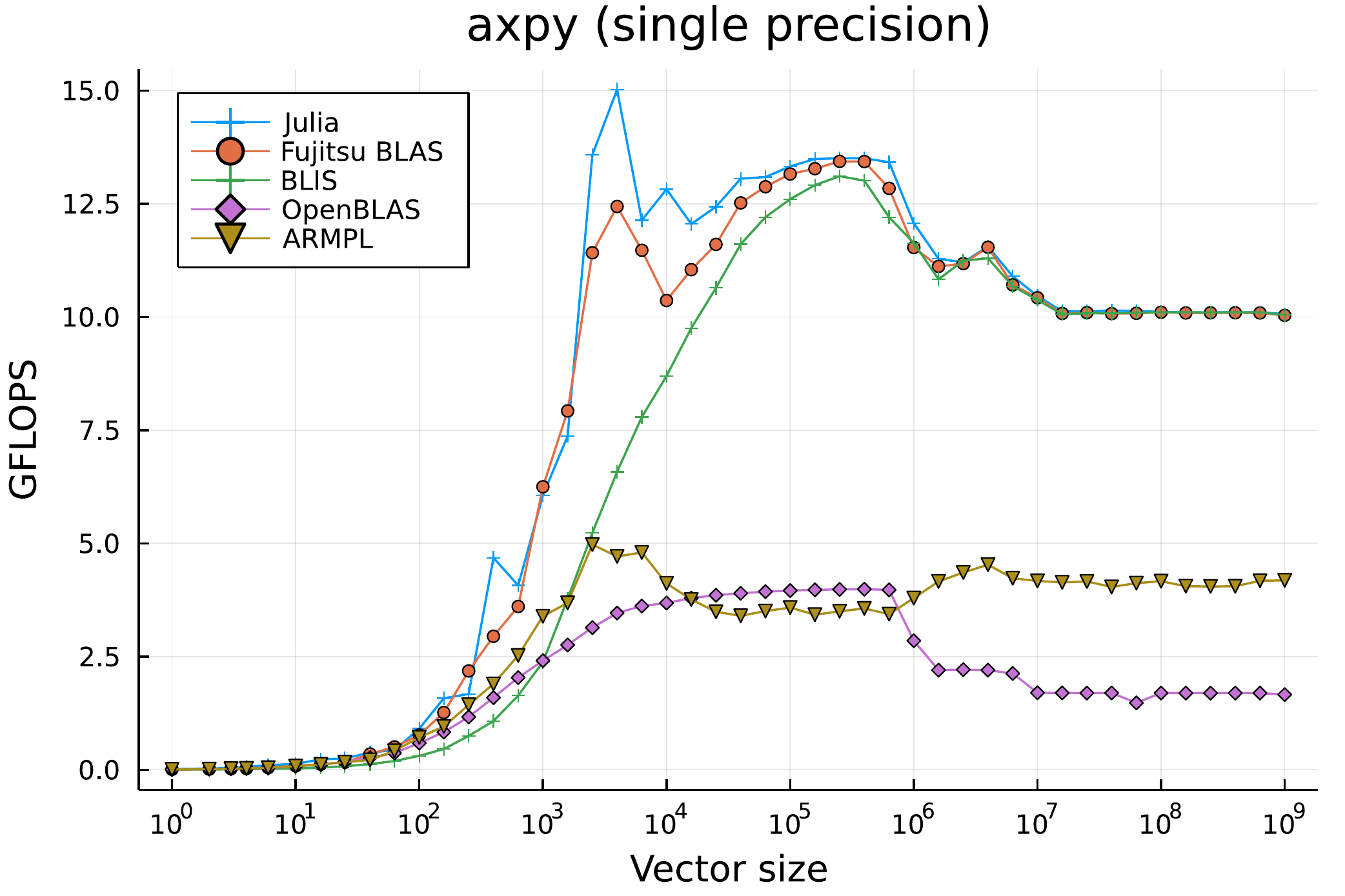} \\[1em]
    \includegraphics[width=\columnwidth]{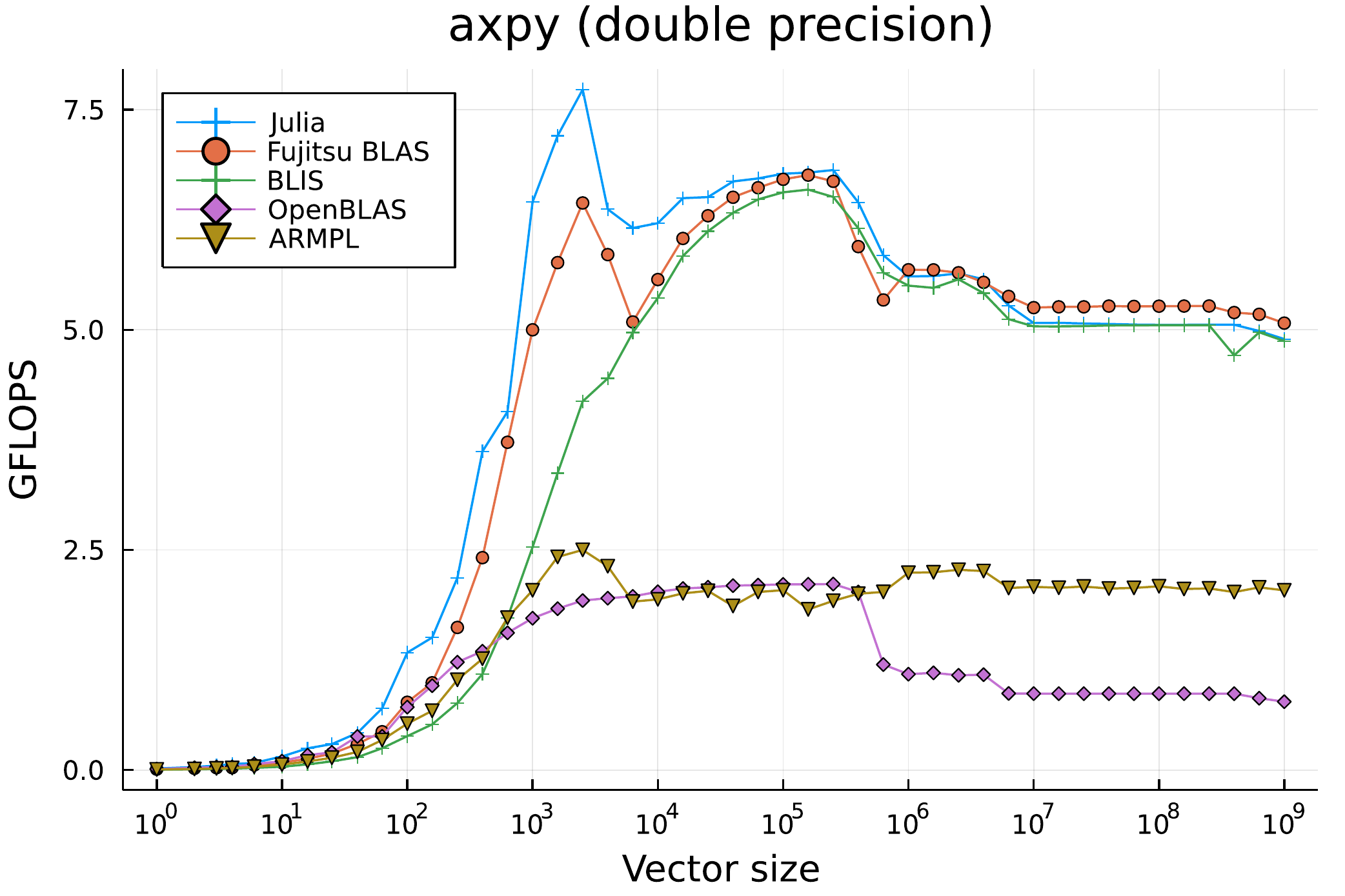} \\[1em]
    \caption{Performance comparison of \texttt{axpy} implementations in Julia versus Fujitsu BLAS, BLIS, OpenBLAS, and ARMPL on Fugaku, using half precision (top panel, \texttt{Float16}), single precision (middle, \texttt{Float32}) and double precision (bottom, \texttt{Flaot64}). A single thread is used in all benchmarks. The vector size refers to the length of the vectors $\pmb{x}$ and $\pmb{y}$ in \cref{eq:axpy}. GFLOPS are gigaFLOPS, the number floating-point operations per second executed by each program.  The panel for half precision shows performance only for Julia, because half-precision implementations of \texttt{axpy} are not available for the other binary libraries (Fujitsu BLAS, BLIS, OpenBLAS, ARMPL).}
    \label{fig:axpy}
\end{figure}
We compare performance of the above generic Julia function, using Julia v1.7.2 (based on LLVM 12) having set the environment variable \texttt{JULIA\_LLVM\_ARGS} to \texttt{-aarch64-sve-vector-bits-min=512}, with the following binary libraries:
\begin{itemize}
    \item vendor's Fujitsu BLAS, from module \texttt{lang/tcsds-1.2.35} on Fugaku, library called \texttt{libfjlapackexsve\_ilp64.so}, with ILP64 and support for SVE instructions,
    \item BLIS version 0.9.0,
    \item OpenBLAS version 0.3.20, built with the Spack package manager version 0.19, using the specification \texttt{openblas@0.3.20 \%gcc@8.5.0 +ilp64 symbol\_suffix=64\_} (building OpenBLAS with the Fujitsu compiler resulted in multiple compilation errors\footnote{See \url{https://github.com/xianyi/OpenBLAS/issues/3692} and \url{https://github.com/spack/spack/issues/31675}.}),
    \item ARM Performance Libraries (ARMPL) version 22.0.2 for RHEL 8 with GCC 8.2, library called \texttt{libarmpl\_ilp64.so}, single-threaded, with ILP64.
\end{itemize}
For these benchmarks we use \texttt{libblastrampoline}\footnote{\url{https://github.com/JuliaLinearAlgebra/libblastrampoline}.}, a library which uses Procedure Linkage Table (PLT) trampolines to forward BLAS calls to a chosen library (e.g., Fujitsu BLAS or BLIS) at runtime with near-zero overhead compared to the complexity of the routines invoked, without having to recompile an application to link to a different BLAS library.
\Cref{fig:axpy} shows the results of these benchmarks.
Both OpenBLAS and ARMPL show poor performance for this routine, likely because it is not taking full advantage of A64FX vectorization capabilities.
We note that there are no implementations of \texttt{axpy} for half-precision floating-point numbers in Fujitsu BLAS, BLIS, OpenBLAS, and ARMP, whereas Julia is able to generate code for the type-generic function \texttt{axpy!} with half-precision Float16 numbers.
Also, Julia's implementation consistently outperforms BLIS, OpenBLAS and ARMPL implementations of \texttt{axpy} in both single and double precision, it is competitive with Fujitsu BLAS across all sizes, and it achieves the best peak performance in all cases.

Thanks to contribution of ARM engineers, auto-vectorization in LLVM 14 is able to target SVE/SVE2 by default when available~\cite{arm:llvm14}.
Preliminary benchmarks performed with the development version of Julia v1.9, based on LLVM 14.0.2, showed similar results to those reported in \Cref{fig:axpy} and the generated LLVM IR for the \texttt{axpy!} function uses \texttt{llvm.vscale} instrinsics, but without having to
set the environment variable \texttt{JULIA\_LLVM\_ARGS} to \texttt{-aarch64-sve-vector-bits-min=512}, which improves the ability of writing high-performance code in Julia.

\subsubsection{Distributed computing with \texttt{MPI.jl}}

MPI is a communication protocol for distributed programs which run on multiple cores and is a staple in the HPC field: it is the de-facto standard for communication in highly parallel applications.
\texttt{MPI.jl} is a Julia package to interface with this protocol~\cite{Byrne2021}.
Studies about MPI communications with \texttt{MPI.jl} on x86 architecture were conducted by~\cite{hunold:julia-mpi-benchmarking,rizvi:communication-intensive-julia}, who found relatively little overhead on AMD and Intel CPUs.

R-CCS presented at the \emph{7th meeting for application code tuning on A64FX computer systems} the results of performance benchmarks of MPI communication on Fugaku~\cite{nakamura:fugaku-mpi}.
These benchmarks were run with the \emph{Intel MPI Benchmarks}\footnote{\url{https://github.com/intel/mpi-benchmarks}.} (IMB) suite with the Fujitsu MPI library, based on Open MPI and optimised for TofuD.
In order to measure performance of MPI communications on A64FX using \texttt{MPI.jl}, we developed \texttt{MPIBenchmarks.jl}\footnote{\url{https://github.com/JuliaParallel/MPIBenchmarks.jl}.}, a package to run Julia benchmarks comparable to some of those in the IMB suite.

\begin{figure}
    \centering
    \includegraphics[width=\columnwidth]{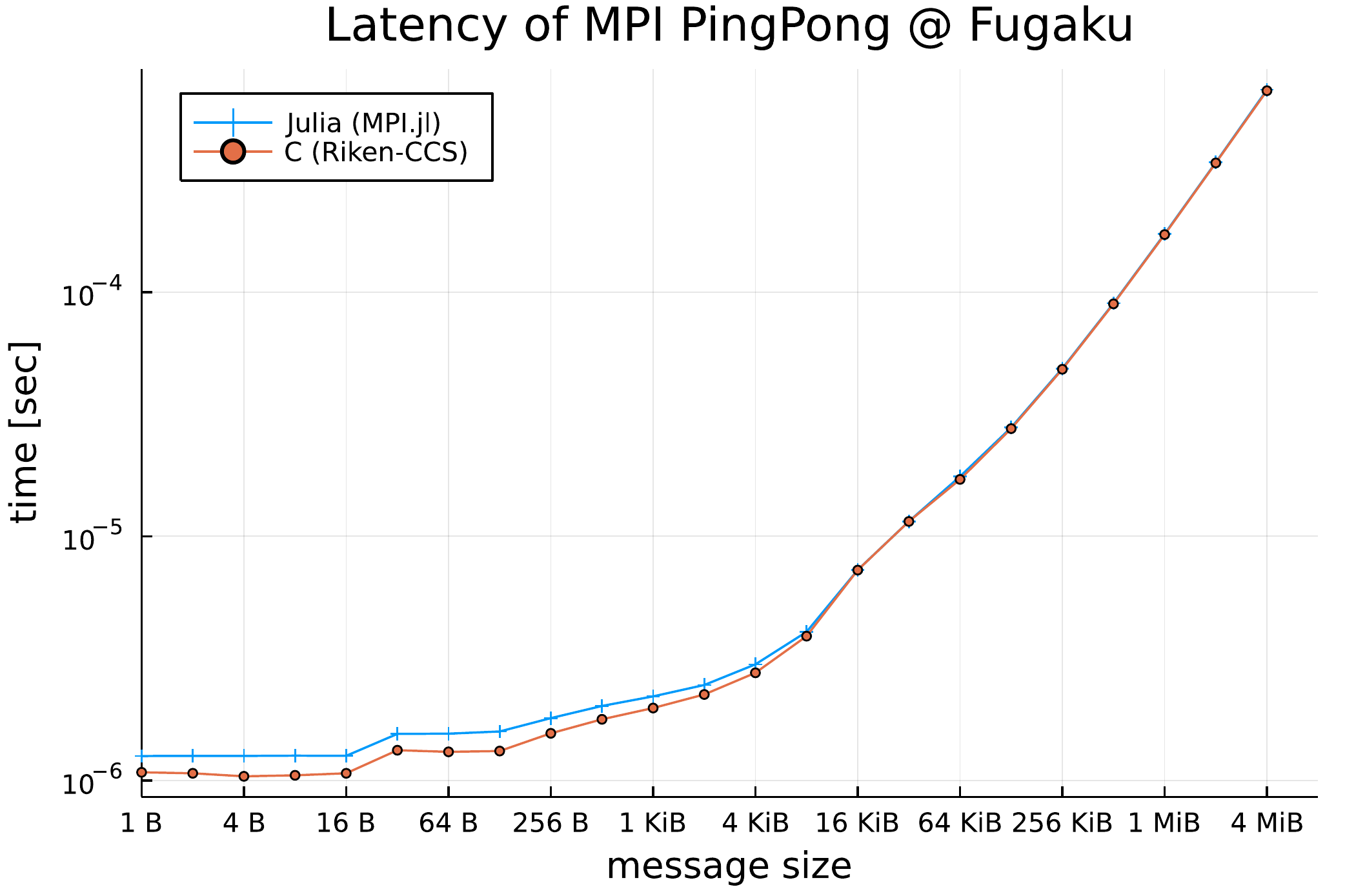} \\[1em] \includegraphics[width=\columnwidth]{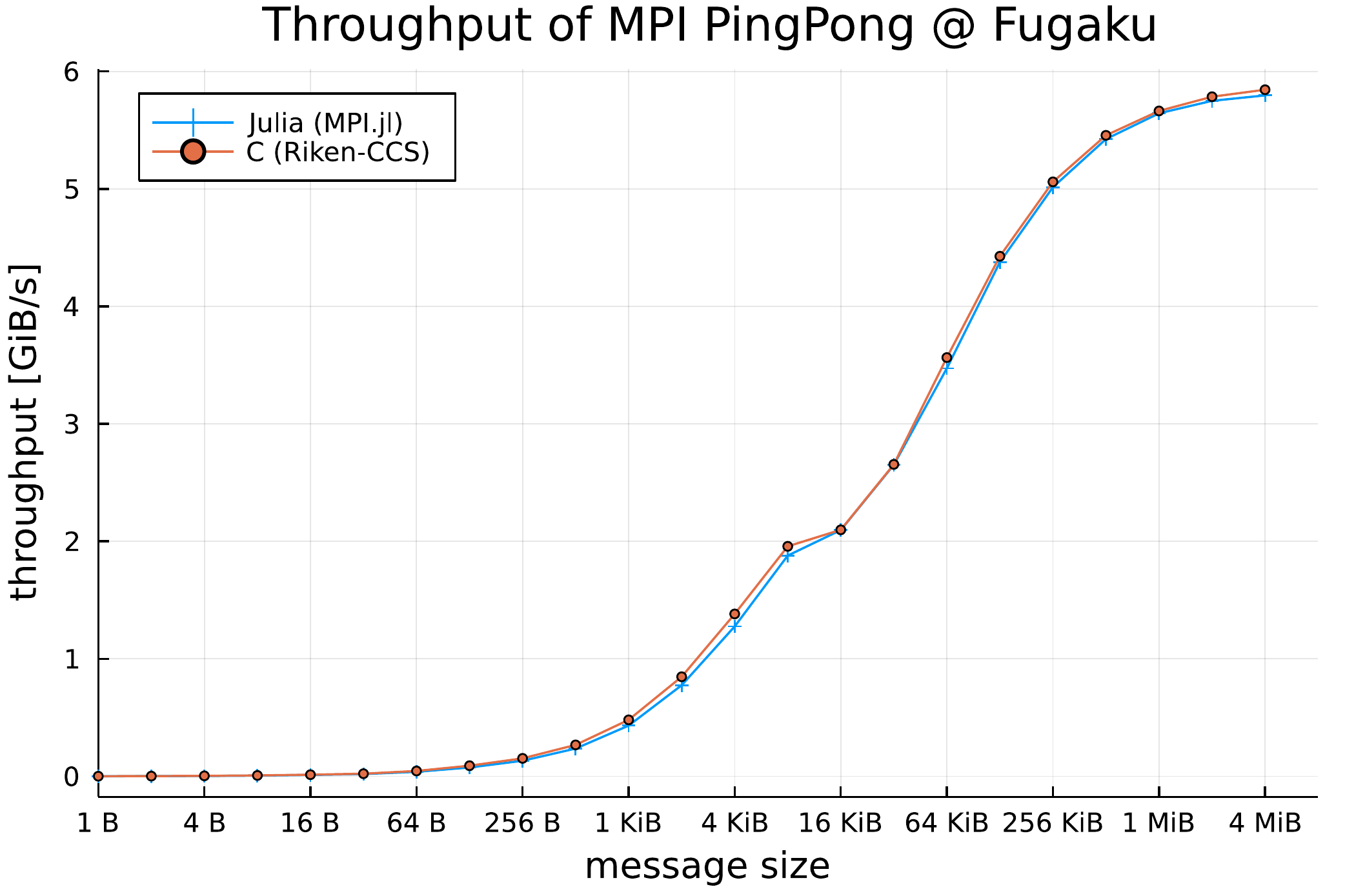}
    \caption{Comparison of latency (top panel) and throughput (bottom panel) of inter-node point-to-point MPI communication between using \texttt{MPI.jl} in Julia and IMB benchmarks in C (results provided by R-CCS in~\cite{nakamura:fugaku-mpi}). Fugaku scheduler setup: \texttt{-L "node=2" --mpi "max-proc-per-node=1"}.}
    \label{fig:mpi-p2p}
\end{figure}
\begin{figure}
    \centering
    \includegraphics[width=\columnwidth]{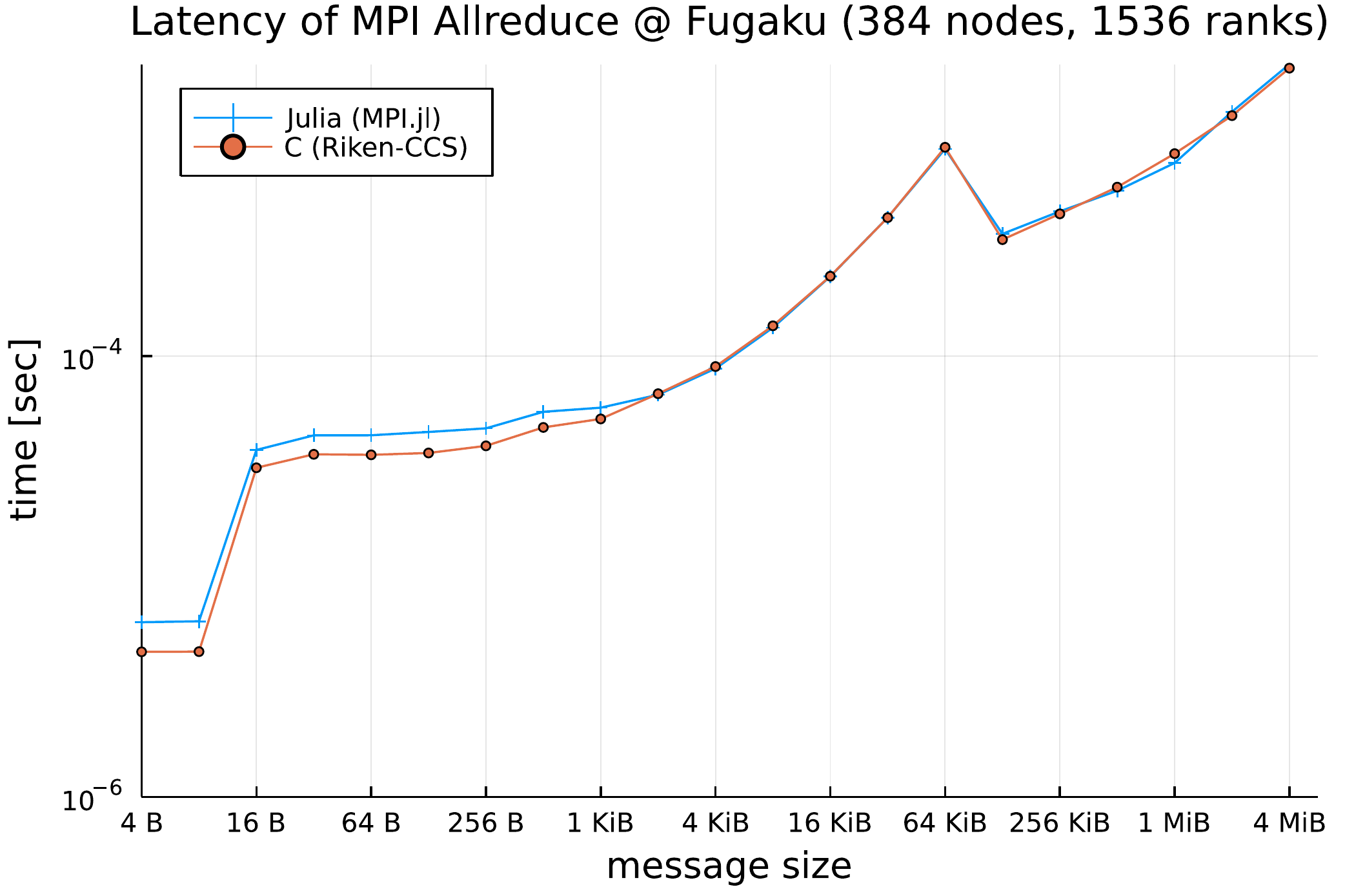} \\[1em]
    \includegraphics[width=\columnwidth]{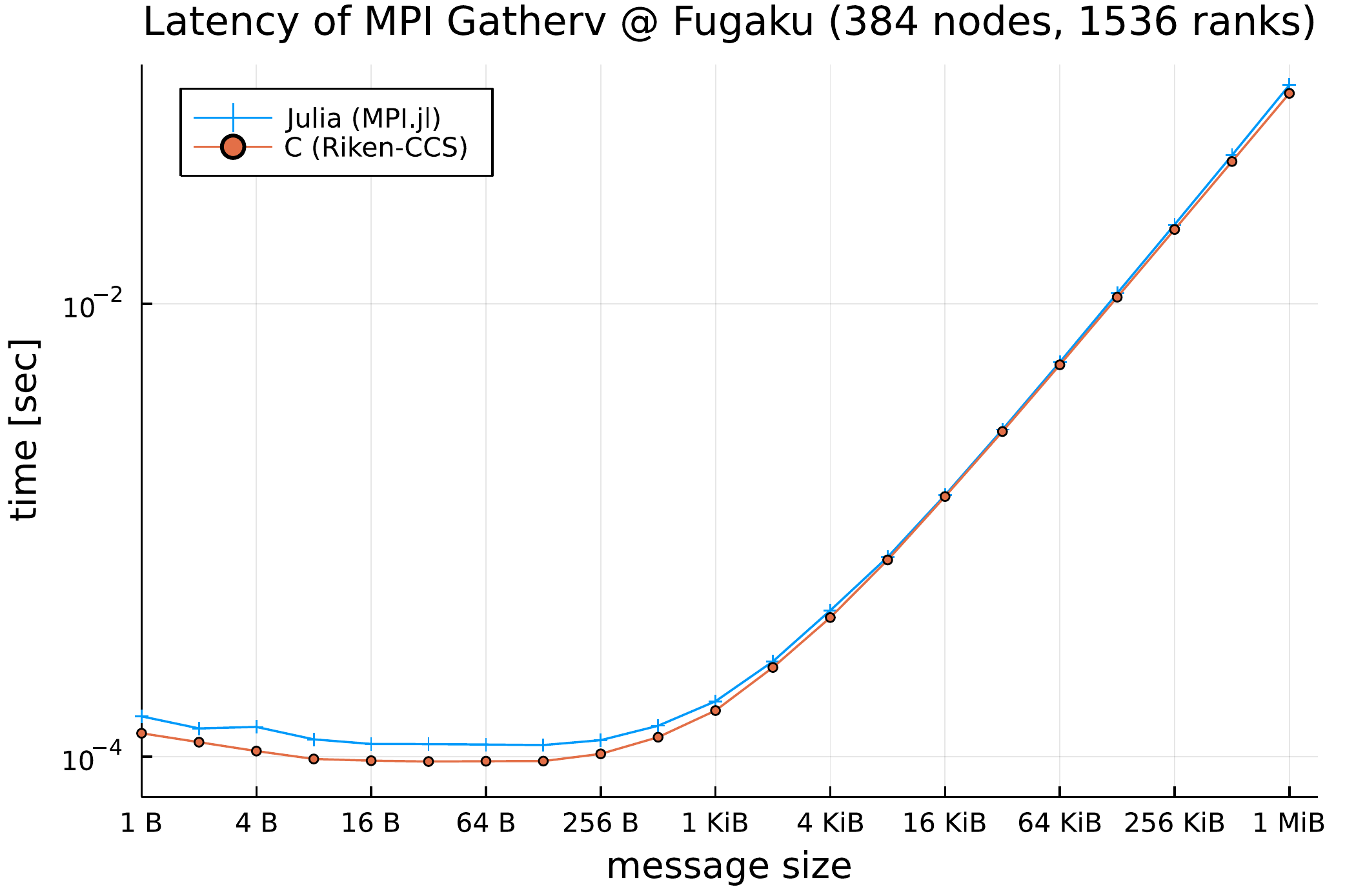} \\[1em]
    \includegraphics[width=\columnwidth]{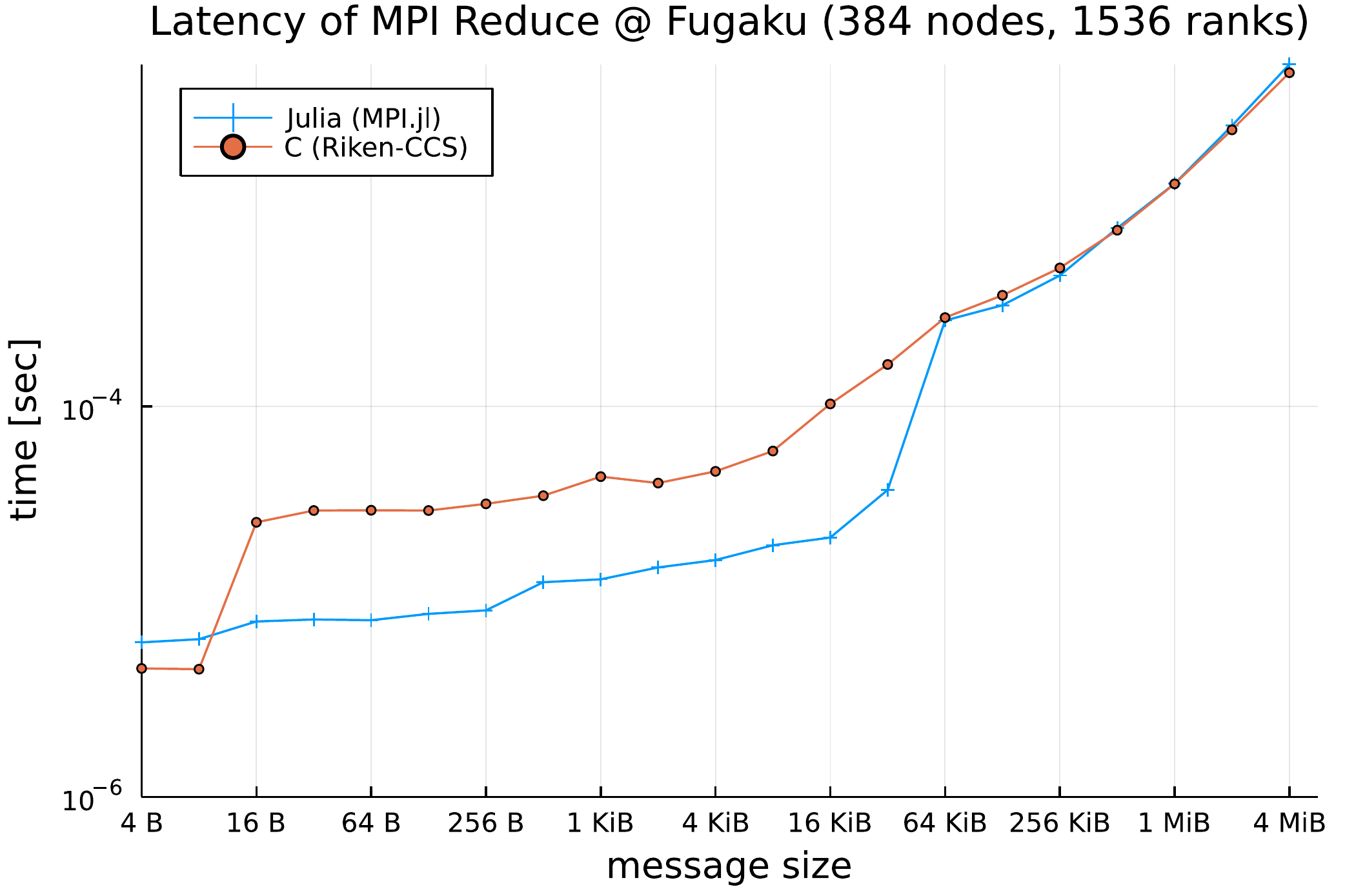}
    \caption{Comparison of latency of collective MPI operations between using \texttt{MPI.jl} in Julia and IMB benchmarks in C (results provided by R-CCS in~\cite{nakamura:fugaku-mpi}): MPI Allreduce (top panel), MPI Gatherv (middle panel), MPI Reduce (bottom panel). Fugaku scheduler setup: \texttt{-L "node=4x6x16:torus:strict-io" -L "rscgrp=small-torus" --mpi proc=1536}.}
    \label{fig:mpi-collective}
\end{figure}
\Cref{fig:mpi-p2p} shows the results of the benchmarks  for a point-to-point operation  (ping-pong), and \Cref{fig:mpi-collective} refers to collective operations (\texttt{Allreduce}, \texttt{Gatherv}, \texttt{Reduce}).
We ran the benchmarks with Julia v1.7.2, using \texttt{MPI.jl} v0.20, which was configured to use the Fujitsu MPI library available in the system.
Point-to-point benchmarks were run with two MPI ranks on two nodes, collective benchmarks were run with 1536 MPI ranks across 384 nodes using the torus layout, to match the scheduler configuration of the R-CCS benchmarks.
\texttt{MPI.jl} typically showed very small overhead for messages larger than 1-2~KiB---peak throughput of ping-pong communication with \texttt{MPI.jl} is within 1\% of that reported by R-CCS---, but slightly larger overhead for messages of smaller sizes.
We note that, contrary to IMB, at the present time \texttt{MPIBenchmarks.jl} does not implement a cache-avoidance mechanism, which may explain why \texttt{MPI.jl} appears to show better latency than IMB for messages with size up to 64~KiB, which corresponds to the size of the L1 cache of the A64FX CPU.
We also observe that, contrary to \cite{hunold:julia-mpi-benchmarking}, we did not find a significant performance drop for the \texttt{Allreduce} operation for larger message sizes.

\subsection{Type flexibility and reduced-precision with \texttt{Float16}}
\label{subsec:float16}

Developing complex applications using \texttt{Float16} is not easy.
On A64FX, even the occasional occurrence of subnormals of \texttt{Float16} ($6\cdot10^{-8}$ to $6\cdot10^{-5}$) causes a heavy performance penalty but a compiler-flag is set to flush them to zero instead\footnote{\url{https://github.com/JuliaLang/julia/issues/40151}.}
The available normal range of \texttt{Float16}, $6\cdot10^{-5}$ to $65,504$, is less than 10 orders of magnitude and scaling is often required to guarantee no under or overflow.
While developing an application that is resilient to the limitations of \texttt{Float16} it is therefore beneficial to retain compatibility to higher-precision formats.
In practice, many complex applications will have parts that are performance-critical, other parts that are precision or range-critical.
An approach is therefore needed that combines a flexibility with the number format for development and performance when ported to various hardware without sacrificing productivity.
Many existing libraries hardcode the number format and deliver performance by essentially duplicating conceptually identical code, which harms productivity and portability to number formats available on modern hardware.
On the other hand, Julia's multiple-dispatch allows the development of fully type-flexible applications, such that the number format, or combinations of different formats, can be chosen at compile time as described in the previous section.
In the following we present one application that prototypes this concept for weather and climate models and runs in \texttt{Float16} on A64FX or in \texttt{Float64} on x86 without changes.

\texttt{ShallowWaters.jl}, a fluid circulation model that solves the shallow water equations for idealized weather and climate simulations, has been developed with a focus on number format-flexibility\cite{Klower2022}.
It runs in \texttt{Float64}, \texttt{Float32}, or \texttt{Float16} alike and in general supports any (custom) number format as long as a standard set of arithmetic operations are implemented.
Functions are written for element types \texttt{T<:AbstractFloat} and Julia dynamically dispatches an arithmetic operations like addition to \texttt{+(x::T,y::T)}, i.e. the respective method defined for the number format \texttt{T}.
While this method can be defined in Julia's Base library (as is the case for floats) any custom number format can be defined by implementing a standard set
of arithmetic operations.
What this set of operations needs to contain depends on the application.
\texttt{ShallowWaters.jl}, for example, only uses transcendental functions like \texttt{log,exp,sin,cos} etc. for precalculating constants and not in the performance-critical main loop.
To optimize the range of numbers occurring \texttt{ShallowWaters.jl}, we developed the analysis-number format \texttt{Sherlogs.jl}, which records a histogram of numbers during the simulation that allowed us to monitor, for example, how a multiplicative scaling $s$ of the equations avoids \texttt{Float16} subnormals. For development purposes we therefore run \texttt{ShallowWaters.jl} with \texttt{T=Sherlog32} (\texttt{Sherlogs.jl}'s equivalent of \texttt{Float32}), and, after choosing $s$, we execute the same code with \texttt{T=Float16,s=s} for performance. For more details see \cite{Klower2022}.
Type-flexibility therefore is not just important for performance code, but can also greatly assist in the development of low precision-resilient code and retains compatibility with higher precision formats.

\begin{figure}
    \centering
    \includegraphics[width=1\columnwidth]{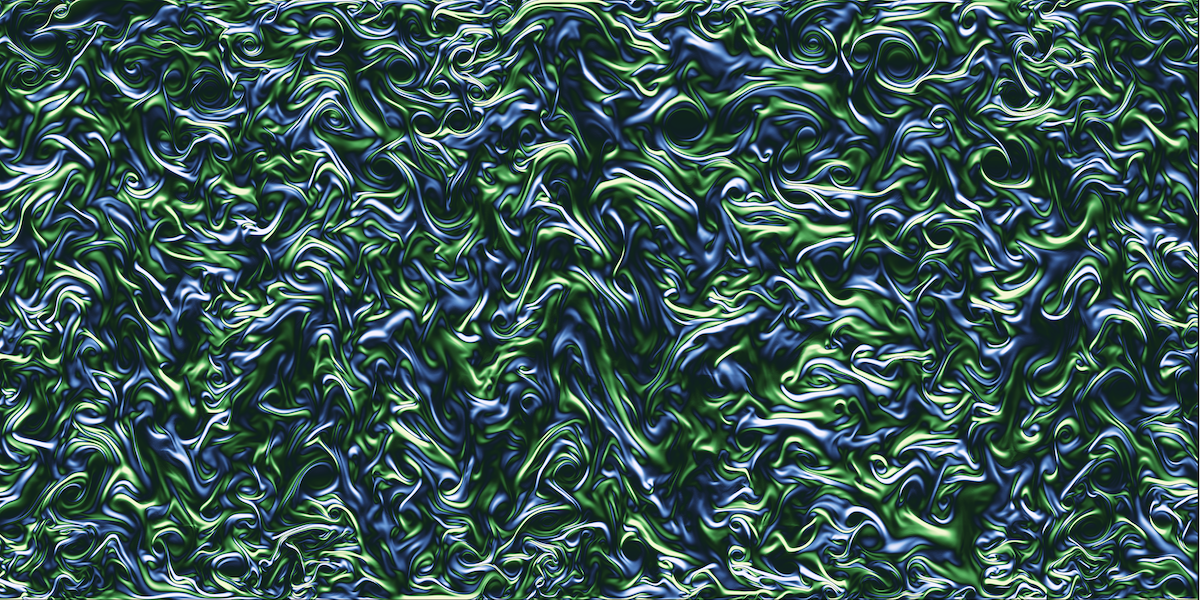}
    \caption{Geophysical turbulence simulated by \texttt{ShallowWaters.jl} using \texttt{Float16} arithmetic on A64FX and a spatial resolution of $3000x1500$ grid points. \texttt{ShallowWaters.jl} uses an identical code base that is dynamically dispatched to any number format. To reduce rounding errors in the precision-critical time integration, \texttt{Float16} simulations includes a compensated summation, which is not required for higher-precision formats. The equivalent \texttt{Float64} simulation is qualitatively indistinguishable but ran $3.6x$ slower \cite{Klower2022}.}
    \label{fig:shallowwaters}
\end{figure}

\begin{figure}
    \centering
    \includegraphics[width=1\columnwidth]{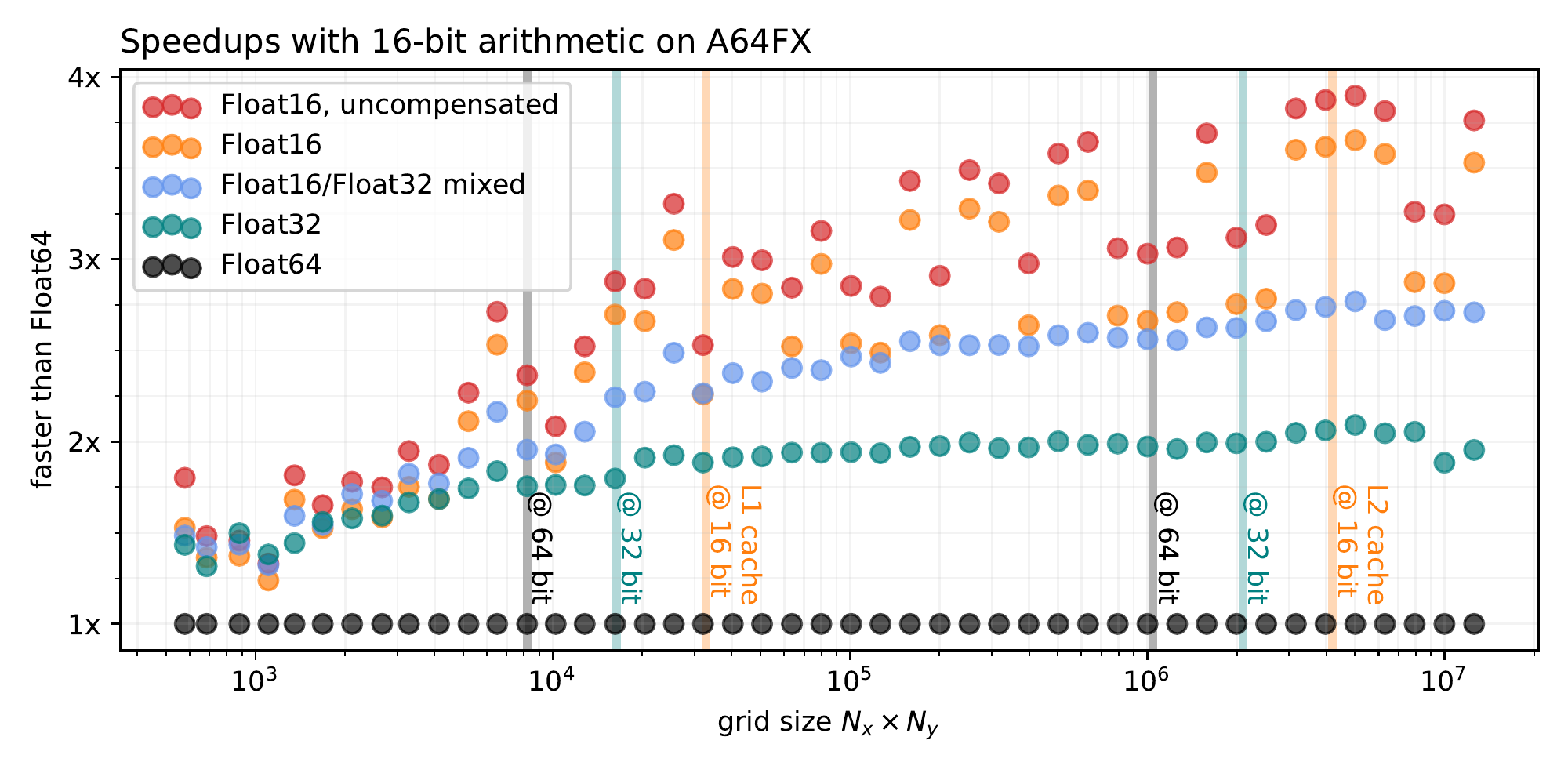}
    \caption{Speedups of low-precision simulations on A64FX with \texttt{ShallowWaters.jl} over \texttt{Float64} with varying problem sizes. \texttt{Float16} has by default a compensated time integration in \texttt{ShallowWaters.jl}, to reduce rounding errors, which causes an about $5\%$ overhead in runtime. \texttt{Float16/32} is a mixed-precision simulation that uses \texttt{Float32} precision for the time integration. Reproduced from \cite{Klower2022}.}
    \label{fig:swm_performance}
\end{figure}

The \texttt{ShallowWaters.jl} simulations with \texttt{Float16} are qualitatively indistinguishable from simulations with \texttt{Float64} (\Cref{fig:shallowwaters}) and rounding errors remain smaller than model or discretization errors.
The precision-critical part is the time integration for which we include a compensated summation that compensates for the rounding error of the previous time step by adding a correction to the next time step.
This introduces a $5\%$ overhead in runtime and therefore clearly outperforms a mixed-precision approach whereby the precision-critical time integration is computed using Float32 (\Cref{fig:swm_performance}).
As \texttt{ShallowWaters.jl} is a memory-bound application it benefits from \texttt{Float16} on A64FX even without vectorization and approaches $4x$ speedups over \texttt{Float64} for large problems ($3000x1500$ grid points, corresponding approximately to array sizes).
\texttt{Float32} simulations are $2x$ faster than \texttt{Float64} over a much wider range of problem sizes.

\section{Opportunities for improvements}
\subsection{General performance}

An evaluation of performance portability of Julia code across multiple architectures, including A64FX, was carried out by~\cite{lin:julia-performance}, who showed that Julia could achieve on this platform performance close to that of equivalent code written in C/C++.
The authors of this study noted that the performance improved sensibly when moving from Julia v1.6, which is based on LLVM 11, to Julia v1.7, based on LLVM 12.
Analogously, we found that the performance of a simple Julia implementation of the BLAS \texttt{axpy} routine can be competitive with that of a vendor implementation of BLAS (Fujitsu BLAS), or another highly optimised library (BLIS).
Furthermore we observed that thanks to recent advancements in LLVM, Julia v1.9 will be able to more easily vectorize the code, without requiring users to manually set LLVM flags.
Thus, owing to the fact it relies on a compiler infrastructure which receives active support from hardware vendors and HPC engineers, Julia can enjoy improvements versions after versions also on very specialised CPUs, despite the fact that to date Julia itself has not received directly any specific optimisation for A64FX.

A64FX is a non-general-purpose CPU, with strong focus on vectorization.
This results in poor performance in some tasks, such as compilation of software.
On Fugaku this issue can be limited by cross-compiling static software optimised for A64FX on the Intel login nodes.
However, Julia is Just-In-Time-compiled (JIT), thus paying the cost of longer compile times in every session whenever a new method needs to be compiled.
Julia currently does not support cross-compilation of code for a different architecture, but there are tools to enable basic ahead-of-time compilation, to generate a system image to reduce the need to compile methods at runtime\footnote{\url{https://github.com/JuliaLang/PackageCompiler.jl}.}.
Improvements in this area of the Julia ecosystem can enhance the ability to run large-scale applications on A64FX and other similar non-general-purpose CPUs.

\subsection{Custom reduction operators in \texttt{MPI.jl}}

An issue that is limiting the ability to run some MPI applications on ARM CPUs is the impossibility to use custom MPI reduction operations on non-Intel architectures due to how they are implemented in \texttt{MPI.jl}\footnote{\url{https://github.com/JuliaParallel/MPI.jl/issues/404}.}.
When this bug will be resolved, it will be possible to run a larger class of Julia MPI programs on ARM systems, including Fugaku.

\subsection{Improved compiler support}
\label{subsec:better_compiler}
As noted in \cref{sec:fp16_support} compilers need be careful in how they handle \texttt{Float16} on platforms that have only software support for it. If they allow for extending precision intermediately, this can cause issues with code as simple as:
\begin{jllisting}
muladd(x, y, z) = x*y+z
\end{jllisting}

which Julia lowers to the following LLVM Intermediate Representation (IR):
\begin{lstlisting}
define half @julia_muladd(half %0, half %1, half %2) {
top:
  %3 = fmul half %0, %1
  %4 = fadd half %3, %2
  ret half %4
}
\end{lstlisting}

In order to ensure the consistency between hardware and software, Julia inserts \texttt{fpext} and \texttt{fptrunc} operations explicitly:

\begin{lstlisting}
define half @julia_muladd(half %0, half %1, half %2){
top:
  %3 = fpext half %0 to float
  %4 = fpext half %1 to float
  %5 = fmul float %3, %4
  %6 = fptrunc float %5 to half
  %7 = fpext half %6 to float
  %8 = fpext half %2 to float
  %9 = fadd float %7, %8
  %10 = fptrunc float %9 to half
  ret half %10
}
\end{lstlisting}

On systems with full hardware support this is clearly sub-optimal and there is ongoing work in LLVM and Julia to address this issue. In particular we will need to extend Julia's multi-versioning support to detect full \texttt{Float16} hardware support and then selecting a copy of the cache code that was compiled without inserting these extra conversion operations.

\section{Conclusions}

In this paper we evaluated the use of the Julia programming language on A64FX, in particular with regards to the ability to easily generate efficient code on a non-general purpose CPU; the possibility of writing type-generic code which can take advantage of hardware acceleration; and the overhead of running distributed applications with \texttt{MPI.jl} on a large supercomputer.

With the example of a simple Level 1 BLAS routine, we found that, by leveraging the work done in the LLVM compiler, Julia allows users to write high-level generic code which is compiled down to native code for A64FX with performance competitive with that of specialised libraries.
Recent improvements in LLVM will further enhance the ability to automatically vectorize the code and fully use SVE instructions, particularly important with the introduction of more ARM CPUs supporting this extension, such as Neoverse V1 and N2.

The \texttt{MPI.jl} package provides a natural interface to the MPI protocol, which allows calling directly MPI libraries optimised for the network of the current system.
Communication benchmarks of \texttt{MPI.jl} showed relatively little overhead compared to the performance analysis carried out by R-CCS, especially for sufficiently large messages, in line with the findings of \cite{hunold:julia-mpi-benchmarking,rizvi:communication-intensive-julia}: peak throughput of point-to-point communications was nearly identical to that measured using the IMB suite in C.

We then presented the case of the \texttt{ShallowWaters.jl} package, a fully type-flexible fluid circulation solver for the shallow water equations which, thanks to Julia's multiple dispatch and code generation, can run on different architectures (e.g., x86 and ARM) and with different types (e.g., \texttt{Float16} and \texttt{Float64}) by using different data types as input to the program, without changing the code base for the different situations.
This showed that Julia is particularly well suited for developing generic numerical code which can effortlessly use different numerical data types without sacrificing performance, boosting scientific productivity.
We were also able to verify that running simulations with \texttt{Float16} on A64FX delivers a nearly $4x$ speedup over \texttt{Float64}.

However, some challenges in the use of Julia on A64FX remain.
Prior to Julia v1.9, the now in-development version which will be based on LLVM 14, SVE instructions are enabled only when manually setting the LLVM flag \texttt{-aarch64-sve-vector-bits-min=512}, which however also causes crashes in many situations\footnote{See for example \url{https://github.com/JuliaLang/julia/issues/43069}, \url{https://github.com/JuliaLang/julia/issues/44263}, \url{https://github.com/JuliaLang/julia/issues/44401}.}.
\texttt{Float16} numbers are promoted to \texttt{Float32} numbers even when hardware support for 16-bit numbers is available, although it is possible to compile Julia with this mechanism disabled (as done here in section \ref{subsec:float16}).
Compilation latency on A64FX can sometimes hinder runtime performance of Julia, more than on general-purpose CPUs, in particular in short-running tasks.
\texttt{MPI.jl} does not currently support custom MPI reduction operators on ARM CPUs, which poses a limitation on running certain distributed application in Julia on these systems.
While more work is needed to fully unlock Julia's potential on A64FX-based clusters---and a large part of it is on-going both in LLVM and Julia itself and its ecosystem, not limited to this CPU---, we found that this is already an effective language for writing generic and high-performance code for A64FX.

\section*{Acknowledgment}
The authors thank Tim Besard, Jameson Nash and Simon Byrne for their inputs and work on both Julia and \texttt{MPI.jl}.
This work used computational resources of the supercomputer Fugaku provided by RIKEN through the HPCI System Research Project (Project ID:~ra000019).
The ShallowWaters.jl simulations were run on the Isambard UK National Tier-2 HPC Service operated by GW4 and the UK Met Office, and funded by the Engineering and Physical Sciences Research Council EPSRC (EP/P020224/1).
MK gratefully acknowledges funding from the European Research Council under the European Union’s Horizon 2020 research an innovation programme (grant no.~741112).
VC gratefully acknowledges funding from NSF (grants OAC-1835443, OAC-2103804, AGS-1835860, and AGS-1835881), DARPA under agreement number HR0011-20-9-0016 (PaPPa). This research was made possible by the generosity of Eric and Wendy Schmidt by recommendation of the Schmidt Futures program, by the Paul G. Allen Family Foundation, Charles Trimble, Audi Environmental Foundation. This material is based upon work supported by the Department of Energy, National Nuclear Security Administration under Award Number DE-NA0003965. The views and opinions of authors expressed herein do not necessarily state or reflect those of the United States Government or any agency thereof. The U.S. Government is authorized to reproduce and distribute reprints for Government purposes notwithstanding any copyright notation herein.

\bibliographystyle{IEEEtran}
\bibliography{references,vchuravy}

\end{document}